\documentclass[10pt,journal]{IEEEtran}

\usepackage[pdftex]{graphicx}
\usepackage{algorithm}
\usepackage{color}
\usepackage{caption}
\usepackage{algpseudocode}
\definecolor{ForestGreen}{RGB}{162,52,0}

\usepackage{xcolor}
\usepackage{amsmath}
\usepackage{amssymb}
\usepackage{latexsym}
\usepackage{CJK}
\usepackage{url}
\usepackage{array}
\usepackage{ragged2e}
\usepackage{makecell}
\usepackage{multirow}
\usepackage{threeparttable}

\ifCLASSOPTIONcompsoc
  \usepackage[nocompress]{cite}
\else
  \usepackage{cite}
\fi

\ifCLASSINFOpdf
\else
\fi

\ifCLASSOPTIONcompsoc
  \usepackage[caption=false,font=footnotesize,labelfont=sf,textfont=sf]{subfig}
\else
  \usepackage[caption=false,font=footnotesize]{subfig}
\fi

\begin{document}

\title{Incorporating Distributed DRL into Storage Resource Optimization of Space-Air-Ground Integrated Wireless Communication Network}

\author{Chao Wang, Lei Liu, Chunxiao Jiang,~\IEEEmembership{Senior Member,~IEEE}, Shangguang Wang,~\IEEEmembership{Senior Member,~IEEE}, \\ Peiying Zhang,~\IEEEmembership{Member,~IEEE}, and Shigen Shen,~\IEEEmembership{Member,~IEEE}

\thanks{This work is partially supported by the National Key Research and Development Program of China under Grant 2020YFB1804800, partially supported by the National Natural Science Foundation of China under Grant 61922050 and 62001357, partially supported by the Shandong Provincial Natural Science Foundation, China under Grant ZR2020MF006, partially supported by the Open Foundation of State key Laboratory of Networking and Switching Technology (Beijing University of Posts and Telecommunications) under Grant SKLNST-2021-1-17, partially supported by the Guangdong Basic and Applied Basic Research Foundation under Grant 2020A1515110079, partially part by the China Postdoctoral Science Foundation under Grant 2021M692501, and partially supported by the Graduate Student Innovation Project Funding Project of China University of Petroleum (East China) under Grant YCX2021127. \textit{(Corresponding authors: Lei Liu and Peiying Zhang)}.}
\thanks{Chao Wang is with the College of Computer Science and Technology, China University of Petroleum (East China), Qingdao 266580, China. E-mail: wangch\_upc@qq.com}
\thanks{Lei Liu is with the State Key Laboratory of Integrated Services Networks, Xidian University, Xi'an 710071, China, and also with the Xidian Guangzhou Institute of Technology, Guangzhou 510555, China. E-mail: leiliu@xidian.edu.cn}
\thanks{Chunxiao Jiang is with the Beijing National Research Center for Information Science and Technology, Tsinghua University, Beijing 100084, China, and also with Tsinghua Space Center, Tsinghua University, Beijing 100084, China. E-mail: jchx@tsinghua.edu.cn}
\thanks{Shangguang Wang is with the State Key Laboratory of Networking and Switching Technology, Beijing University of Posts and Telecommunications, Beijing 100876, China. E-mail: sgwang@bupt.edu.cn}
\thanks{Peiying Zhang is with the College of Computer Science and Technology, China University of Petroleum (East China), Qingdao 266580, China, and also with the State Key Laboratory of Networking and Switching Technology, Beijing University of Posts and Telecommunications, Beijing 100876, China. E-mail: zhangpeiying@upc.edu.cn}
\thanks{Shigen Shen is with the Department of Computer Science and Engineering, Shaoxing University, Shaoxing 312000, China. E-mail: shigens@usx.edu.cn}
}

\markboth{IEEE JOURNAL OF SELECTED TOPICS IN SIGNAL PROCESSING,~Vol.~XX, No.~XX, XX~2021}
{}

\maketitle
\begin{abstract}
Space-air-ground integrated network (SAGIN) is a new type of wireless network mode. The effective management of SAGIN resources is a prerequisite for high-reliability communication. However, the storage capacity of space-air network segment is extremely limited. The air servers also do not have sufficient storage resources to centrally accommodate the information uploaded by each edge server. So the problem of how to coordinate the storage resources of SAGIN has arisen. This paper proposes a SAGIN storage resource management algorithm based on distributed deep reinforcement learning (DRL). The resource management process is modeled as a Markov decision model. In each edge physical domain, we extract the network attributes represented by storage resources for the agent to build a training environment, so as to realize the distributed training. In addition, we propose a SAGIN resource management framework based on distributed DRL. Simulation results show that the agent has an ideal training effect. Compared with other algorithms, the resource allocation revenue and user request acceptance rate of the proposed algorithm are increased by about 18.15\% and 8.35\% respectively. Besides, the proposed algorithm has good flexibility in dealing with the changes of resource conditions.
\end{abstract}


\IEEEpeerreviewmaketitle

\section{Introduction}\label{part1}

\begin{figure}[!h]
\centering
\includegraphics[width=1\columnwidth]{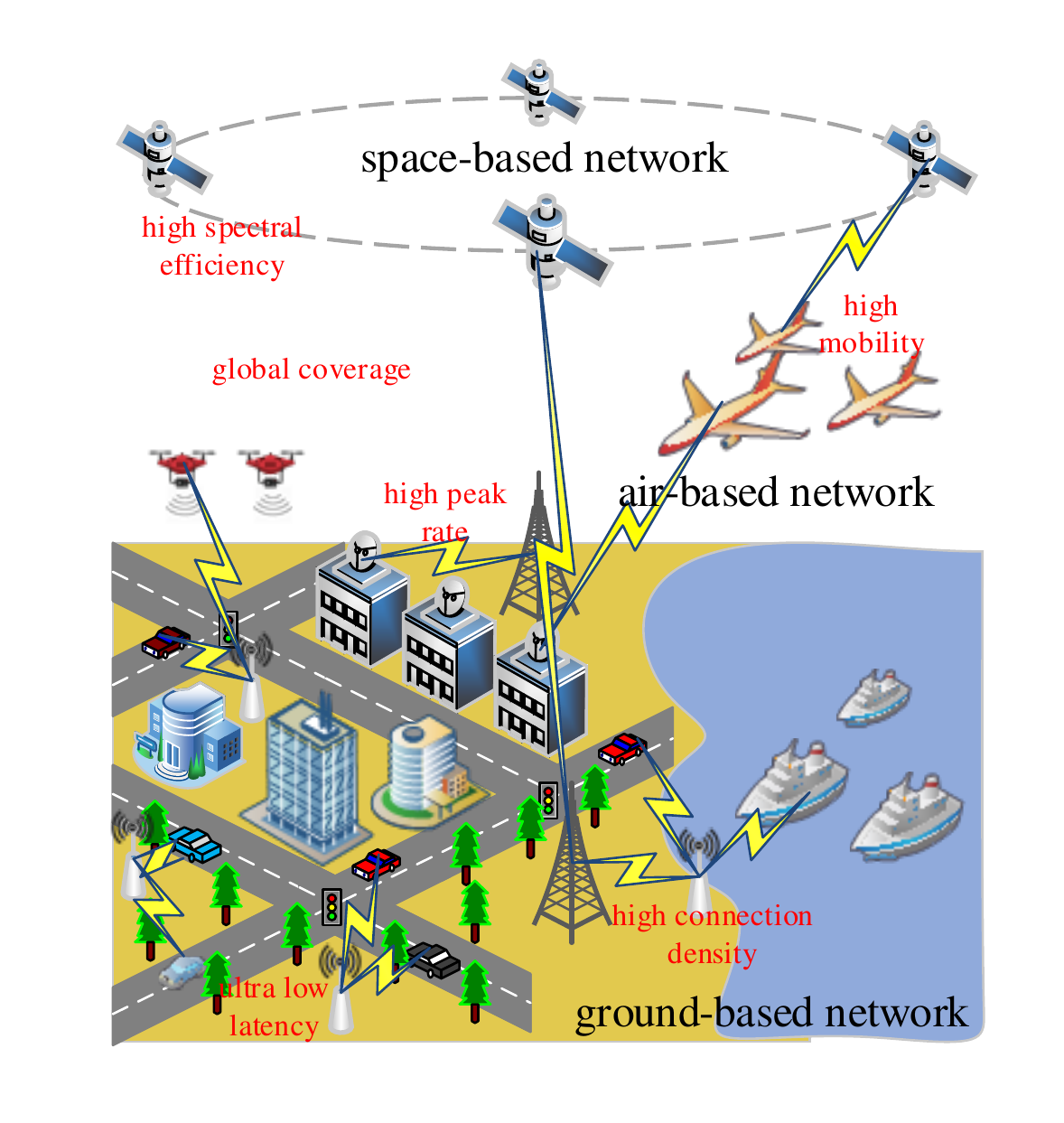}
\caption{Typical space-air-ground integrated network application scenario. The space-air-ground integrated network can effectively cope with the mission requirements such as ultra-low latency transmission, high-density connection, high mobile speed and global coverage in 6G.}
\label{fig_1}
\end{figure}

Artificial intelligence (AI) and wireless communication technology have rapidly spread in recent years. People all over the world have enjoyed the convenience brought by network services \cite{ed1}. Especially in order to achieve the global coverage of wireless networks, a new type of space-air-ground integrated network (SAGIN) has been proposed to provide three-dimensional comprehensive connection services \cite{l1}. SAGIN needs to realize the seamless integration of space-based network, air-based network and ground-based network. Achieving ultra-reliable, low-latency wireless communication is an important goal of SAGIN. Satellites and unmanned aerial vehicles (UAVs) participate in network construction and provide services as air users or base stations (BSs) \cite{j7}. As an emerging wireless network architecture and new service paradigm, SAGIN can provide a reliable, stable and robust service supply for end users worldwide. A typical SAGIN scenario is shown in Fig. \ref{fig_1}. The performance of SAGIN is largely dependent on 6G. SAGIN supported by 6G can bring benefits such as global coverage, ultra-low latency and high connection density, but 6G technology is not yet mature \cite{j6}. Especially for network operators, how to coordinate, manage and schedule the network resources of SAGIN is a severe challenge \cite{q1}. Radio network resource management faces severe challenges, including storage, spectrum, computing resource allocation, and joint allocation of multiple resources \cite{jcx1,jcx2}. With the rapid development of communication networks, the integrated space-ground network has also become a key research object \cite{jcx3}.

Space-based network and air-based network are two special network segments in SAGIN. Satellites and UAVs can act as air servers and air terminal users \cite{j2,ed2}. They are connected to each other, forming a complex space-based network system and space-based network system. Satellites and UAVs are always in motion, so the topological structure of space-based network and air-based network is dynamically changing. In order to ensure the flight speed and reduce the launch cost, the volume of satellites and UAVs are strictly designed, which leads to a problem that the storage performance of air nodes is limited \cite{l2}. SAGIN requires different network segments to have certain computing, storage and communication capabilities, so how to efficiently manage the storage resources of space-based network and air-based network requires in-depth research.

Traditional Internet architecture usually adopts centralized resource management. Centralized system refers to the central server composed of one or more computers. All business units are deployed to the server, and all data storage, management and calculation are carried out on the central server \cite{r3}. Since all services need to be aggregated to a central server for centralized processing, it greatly increases the congestion probability of the network channel. Especially for the SAGIN, long-distance channel transmission may produce higher latency. In recent years, distributed system has become a more popular way of network management \cite{j3}. Distributed system deploys servers on different network computers. The edge server can serve end users in the region, and the system tasks can be completed through message passing coordination between edge servers. Taking into account the characteristics of the extensive deployment of SAGIN, it will be extremely necessary to adopt a distributed management solution for SAGIN \cite{ed3}.

Dynamic and intelligent physical network resource management is one of the core businesses of SAGIN \cite{zz1,j5}. Physical network resource management methods based on heuristic or offline algorithms, such as particle swarm algorithm \cite{e1}, simulated annealing algorithm \cite{e2}, genetic algorithm, etc., have all played an active role in physical network resource allocation \cite{z4,e3}. It cannot be ignored that the above algorithms have many limitations in practical applications. On the one hand, heuristic algorithms are not suitable for large-scale networks, and the final results often fall into local optimal solutions. On the other hand, the above algorithms ignore the time limit of real-world applications. In the case of large-scale network or complex structure, the time cost of algorithm execution increases, which is unacceptable for some real-time applications. Deep reinforcement learning (DRL) is a promising means of network resource management \cite{l3}. SAGIN is a typical high-dimensional space architecture. Because DRL has good perception and decision-making capabilities, it can obtain the best resource management plan through the interaction of intelligent agent with the environment, so it is suitable for high-dimensional space decision issues \cite{r5,z2}.

The management of SAGIN resources based on distributed architecture still faces many practical problems \cite{j4,ed4}. First of all, how to highlight the storage performance characteristics of space-based network and air-based network, and then achieve effective management of storage performance is a problem worthy of consideration. Secondly, SAGIN resource management is a dynamic process, and the network resources of different network segments are heterogeneous. It is necessary to manage the resources in real time while considering the practical requirements such as low cost and high revenue. In order to deal with the above problems, the main work of this paper is summarized as follows.

\begin{enumerate}
\item We model the SAGIN resource allocation problem as a Markov decision process (MDP) and propose a resource management algorithm based on distributed DRL. We use a custom neural network as the intelligent agent. In order to allocate storage resources reasonably, the training environment of agent is constructed by the feature matrix represented by storage resources.
\item We propose a distributed SAGIN resource management architecture. Space-based network, air-based network and ground-based network all adopt distributed management solutions. A DRL model is deployed on each edge server to manage the network resources in the region in real time.
\item We simulate the algorithm from two aspects of training and testing. Compared with other resource management algorithms, it shows that the proposed algorithm has good resource allocation performance. In addition, we verify the flexibility of the algorithm by changing the storage resource requirements requested by end users.
\end{enumerate}

This paper is organized as follows. Section \ref{part2} introduces the related research progress of network resource management. Section \ref{part3} gives the system model and formulates the related problems. Section \ref{part4} introduces the realization of SAGIN storage resources management algorithm based on distributed DRL, and then we give a distributed-based SAGIN resource management architecture. We conduct simulation experiments and analyze the results in section \ref{part5}. Finally, we summarize the full paper.

\section{related work}\label{part2}

This section mainly discusses the research progress of network resource management from three aspects: SAGIN resource management, machine learning (ML)-based network resource management and distributed network resource management.

\subsection{Resource Management in Space-Air-Ground Integrated Network}

The ultimate goal of building SAGIN is to serve end users, so the resource management of SAGIN is carried out under specific application scenarios. Li et al. \cite{1} studied the problem of network resource management of SAGIN service for Internet of Vehicles (IoV) users. Heterogeneous network resource management was defined as the dynamic service function chain (SFC) mapping and scheduling problem. Firstly, the authors established a mathematical model for the migration cost and migration delay of SFC, and then used mixed integer linear programming (MILP) to solve the dynamic virtual network function (VNF) mapping and scheduling process. Finally, the authors solved the MILP problem through two heuristic algorithms based on tabu search. Cao et al. \cite{2} also studied the issue of SAGIN service for IoV. Different from the reference \cite{1}, Cao et al. combined IoV with cloud computing. Based on this, they proposed a SAGIN-IoV optimization model architecture that took into account energy consumption, delay and resource utilization. In order to realize the autonomous decision-making of the edge network, the authors proposed an improved optimization algorithm, which effectively realized the resource scheduling of SAGIN. Since the power of smart devices in the remote Internet of Things (IoT) was limited, reference \cite{3} used UAVs as repeaters to connect the communication between network satellites and ground devices. Specifically, the authors jointly optimized the device connection, UAV trajectory and energy control, and solved the convex optimization problem through an iterative algorithm. The results showed that this method can effectively improve the revenue of resource allocation. In addition, references \cite{4} and \cite{5} also proposed management schemes for SAGIN resources from different perspectives.

\subsection{Network Resource Management Based on Machine Learning}

Network resource management based on ML has become a research hotspot in recent years \cite{e4,e5,e6}, including network resource allocation algorithms based on deep learning (DL), reinforcement learning (RL) and DRL. Rahman et al. \cite{6} studied the resource allocation problem in fog radio access network (FRAN). The current FRAN is only suitable for static communication mode and is not sensitive to network delay. The authors proposed a joint optimization algorithm based on DRL mode selection, resource and power allocation. The DRL controller can autonomously decide whether to execute the computing task locally or offload it to the cloud server, so as to rationally schedule network resources and reduce latency. Xu et al. \cite{7} paid attention to the problem of resource allocation in software defined network (SDN). However, SDN controllers cannot identify applications autonomously, and manual sampling and identification consume a lot of network resources. Therefore, the authors proposed a method based on DL for applications to perceive VNFs. The specific operation was to map the user function request type and requirement to a specific route, which effectively improved the quality of service of the network. Jiang et al. \cite{8} transformed the joint resource optimization problem in the cellular network into a MILP problem. The authors proposed a resource allocation and power control algorithm based on Q-learning to reduce the high computational complexity of the MILP problem. Experimental results showed that the algorithm can effectively improve throughput and energy efficiency.

5G IoT is an important foundation for the realization of smart city, smart medical and smart vehicles \cite{z3}. However, there are still serious challenges in the effective use of a variety of 5G resources, reducing the cost of resource allocation and privacy protection. The authors of reference \cite{9} proposed a super dense edge computing framework based on AI and blockchain. In this framework, the authors implemented a two period DRL algorithm, which was mainly used to reduce network latency and improve resource utilization. In addition, the privacy of edge devices was effectively guaranteed by federated learning (FL) based on blockchain. Other representative network resource management solutions based on ML include the DL-based resource and deployment framework proposed in \cite{10}.

\subsection{Distributed Network Resource Management}

Compared with centralized management, distributed resource management has a broader application market, especially in the intelligent scenarios such as cloud computing and edge computing. Allybokus et al. \cite{11} studied the problem of resource allocation in distributed SDN systems and proposed a resource allocation algorithm based on the alternating direction multiplier method. Based on this algorithm, a series of fair resource allocation schemes can be generated. Also in SDN network, Pan et al. \cite{12} proposed a distributed resource allocation algorithm for 5G cellular network. The algorithm used the method of maximizing the weighted utility to realize the simultaneous access of licensed and unlicensed frequency bands, which can alleviate the shortage of spectrum resources and ensure the optimization of resources. Based on the concepts of network function virtualization (NFV) and network slicing in virtual network architecture, reference \cite{13} proposed a resource allocation model in a 5G heterogeneous cloud framework. First, the authors modeled the resource allocation as a convex optimization problem, and maximized the system utility function by constructing an auction relationship between slice and data center. After that, the authors formulated a new distributed resource allocation scheme based on the principle of resource fairness, and the results proved that the scheme has good convergence.

The above-mentioned research results only focus on certain application scenarios of network resource management, and do not effectively integrate SAGIN background, ML technology and distributed management solutions. Therefore, they lack comprehensiveness. This paper adopts DRL method to conduct a comprehensive research on SAGIN resource management based on distributed architecture.

\section{System Model}\label{part3}

We consider the general application scenario of SAGIN as shown in Fig. \ref{fig_1}. Its main body includes space-based networks (satellite nodes and inter satellite links), air-based networks (UAV nodes and air links) and ground-based networks (ground nodes and ground links). Different network segments can communicate through inter domain links. Therefore, SAGIN can be regarded as a multi-domain physical network. The end user sends resource requests to the SAGIN. The request may require physical resources in different network segments. Therefore, the user request needs to be mapped to physical nodes in different network segments. Our goal is to rationally arrange physical network resources, increase the revenue of resource allocation on the basis of increasing the acceptance rate of end users' requests, and reduce the cost of resource consumption. To this end, we establish mathematical models for SAGIN, resource allocation constraints, and evaluation indicators. TABLE \ref{tab_1} summarizes the main symbols used in this section.

\begin{table}
\centering
\caption{Notations}
\renewcommand\arraystretch{1.5}
\begin{tabular}{p{15mm} p{60mm}}
\hline
\hline
Notation &  Description \\
\hline
$G^P$ & space-air-ground integrated network \\
$N^P$ & the set of physical nodes in SAGIN \\
$L^P$ & the set of physical links in SAGIN \\
$R^P$ & the set of physical network attributes \\
$N^{PS}$ & the set of space nodes (satellite nodes) \\
$N^{PA}$ & the set of air nodes (UAV nodes) \\
$N^{PG}$ & the set of ground nodes \\
$CPU^P$ & CPU resource capacity of the physical nodes \\
$STO^P$ & storage resource capacity of the physical nodes \\
$BW^P$ & bandwidth resource capacity of the physical links \\
$G_k^R$ & $k$-th end user function request \\
$N^R$ & the set of request nodes \\
$L^R$ & the set of request links \\
$R^R$ & the set of end user function request attributes \\
$CPU^R$ & CPU resource requirement of request nodes \\
$STO^R$ & storage resource requirement of request nodes \\
$BW^R$ & bandwidth resource requirement of request links \\
$\lambda$ & binary variable indicating node resource allocation \\
$\mu$ & binary variable indicating link resource allocation \\
\hline
\hline
\end{tabular}
\label{tab_1}
\end{table}

\subsection{Space-Air-Ground Integrated Network Model}

SAGIN is regarded as a large underlying physical network, which is modeled as a weighted undirected graph $G^P=\{N^P,L^P,R^P\}$. $G^P$ represents the entire SAGIN. $N^P$ represents the set of all physical nodes in SAGIN, and $N^P=\{N^{PS},N^{PA},N^{PG}\}$, where $N^{PS}$ represents the set of space nodes, i.e. satellite nodes, $N^{PA}$ represents the set of air nodes, i.e. UAV nodes, and $N^{PG}$ represents the set of ground physical nodes. $n^{ps}$, $n^{pa}$ and $n^{pg}$ are used to refer to a specific satellite node, UAV node and ground node respectively. $L^P$ is the set of all physical links in SAGIN, $l(n_i^p,n_j^p)$ is the physical link between $n_i^p$ and $n_j^p$. $R^P$ represents the resource attribute set of SAGIN, and $R^P=\{CPU^P,STO^P,BW^P\}$, where $CPU^P$ represents the computing resource attribute of physical nodes, $STO^P$ represents the storage resource attribute of physical nodes, and $BW^P$ represents the bandwidth resource attribute of physical links. When a user function request is successfully mapped, part of the physical resources needs to be occupied. When the request leaves, part of the resources occupied by it will be released.

In order to intuitively reflect the network resource allocation, a mathematical model is established for the end user function request, which is also modeled as an undirected weighted graph $G_k^R=\{N^R,L^R,R^R\}$. $G_k^R$ represents the $k$-th user function request, $N^R$ represents the request node set in the user function request, $L^R$ represents the request link set, $R^R$ represents the resource requirement attribute set of the user function request, and $R^R=\{CPU^R,STO^R,BW^R\}$, where $CPU^R$ represents the computing resource requirement of request nodes, $STO^R$ represents the storage resource requirement of request nodes, and $BW^R$ represents the bandwidth resource requirement of request links. Specifically, $n^r$ is used to represent a specific request node, and $l(n_i^r,n_j^r)$ represents the request link between the request node $n_i^r$ and $n_j^r$.

\subsection{Formulation of Network Resource Allocation Problem}

\subsubsection{Constraint Conditions}

End user function requests often require specific physical network resources. The request node may need to be mapped to different network segments of SAGIN, so the candidate network domain attributes need to be set for the request node, which is defined as follows,
\begin{equation}
candi\_n^r=\{0,1,2|0-N^{PS},1-N^{PA},2-N^{PG}\},
\end{equation}
indicates that the candidate network domain of the request node $n^r$ is one of space-based network, air-based network or land-based network. This formula can also express the location constraint of request node mapping, i.e., for a specific request node, it can only be mapped to only one physical domain. Because different request nodes may undertake different sub-functions, connecting multiple request nodes in series can meet the complete functional requirements of users.

For the same end user request, a request node can only be mapped to a single physical node. In order to connect the corresponding request node, the request link may be mapped to one or more physical links. They are defined as follows,
\begin{equation}
\sum\limits_{k=1}^{|N^R|}\lambda_{n_k^r}^{n^p}=1,\,N^R \in G_i^R,
\end{equation}
\begin{equation}
\sum\limits_{k=1}^{|L^R|}\mu_{l_k^r}^{l^p} \geq 1,\,L^R \in G_i^R,
\end{equation}
where $\lambda_{n_k^r}^{n^p}$ and $\mu_{l_k^r}^{l^p}$ are binary variables. If $\lambda_{n_k^r}^{n^p}=1$, it means that the request node $n_k^r$ is mapped to the physical node $n^p$. If the request node $n_k^r$ is not mapped to the physical node $n^p$, $\lambda_{n_k^r}^{n^p}=0$. Similarly, if $\mu_{l_k^r}^{l^p}=1$, it means that the request link $l_k^r$ is mapped to the physical link $l^p$, and if $\mu_{l_k^r}^{l^p}=0$, it means that the request link $l_k^r$ is not mapped to the physical link $l^p$. $|N^R|$ represents the number of request nodes in the end user function request $G_i^R$, and $|L^R|$ represents the number of request links.

In addition to location constraints, end user function requests also need to follow resource constraints rules. On the premise of determining the candidate network domain, the user function request node can only be mapped to the physical node that meets its resource requirements. Specifically, the CPU resource capacity of the target physical node should be no less than the CPU resource demand of the request node. The storage resource capacity of the target physical node should be no less than the storage resource demand of the request node.
\begin{equation}
CPU(n^p) \geq CPU(n_k^r), \, if \, \lambda_{n_k^r}^{n^p}=1, \, k=1,2,...,|N^R|,
\end{equation}
\begin{equation}
STO(n^p) \geq STO(n_k^r), \, if \, \lambda_{n_k^r}^{n^p}=1, \, k=1,2,...,|N^R|.
\end{equation}

The user function request link can only be mapped to the physical link that meets its resource requirements, which is defined as,
\begin{equation}
BW(l^p) \geq BW(l_k^r), \, if \, \mu_{l_k^r}^{l^p}=1, \, k=1,2,...,|L^R|.
\end{equation}

\subsubsection{Performance Indexes}

The ultimate goal of the network resource allocation algorithm is to improve the revenue of network operators. The revenue is determined by the consumption of different types of network resources, and the calculation method is,
\begin{equation}
Revenue(G_k^R)=\sum\limits_{i=1}^{|N^R|}[CPU(n_i^r)+STO(n_i^r)]+\sum\limits_{j=1}^{|L^R|}BW(l_j^r).
\end{equation}

The above formula calculates the revenue brought by the user function request $G_i^R$ to be successfully mapped.

In addition to the revenue of resource allocation, network operators also care about the cost of resource allocation, which is calculated as follows,
\begin{equation}
\begin{aligned}
Cost(G_k^R)=\sum\limits_{i=1}^{|N^R|}[CPU(n_i^r)+STO(n_i^r)] + \\ \sum\limits_{j=1}^{|L^R|}BW(l_j^r) \cdot hops(l_j^r),
\end{aligned}
\end{equation}
where $hops(l_j^r)$ is the number of path hops of the request link $l_j^r$ due to link splitting. Link splitting means that a function request link is mapped to multiple physical links, and each physical link needs to provide the required link resources for the function request link. Thus, the more times the request link is divided, the more bandwidth resources it consumes. Therefore, when allocating bandwidth resources for the request link, path splitting should be avoided as far as possible.

The above two formulas define the revenue and cost of a single end user request for successful mapping. It is necessary to realize that user function requests and network resource allocation are a long-term continuous process, so it is more reasonable to use the long-term average revenue of network resource allocation to measure algorithm performance, defined as,
\begin{equation}
AR=\lim_{T \to \infty}\frac{\sum\limits_{t=0}^{T}Revenue(G^R,t)}{T}.
\end{equation}

The ratio between revenue and cost generated by the allocation of network resources is called the long-term rate of revenue, which is defined as,
\begin{equation}
R/C=\lim_{T \to \infty}\frac{\sum\limits_{t=0}^{T}Revenue(G^R,t)}{\sum\limits_{t=0}^{T}Cost(G^R,t)}.
\end{equation}

Network operators receive as many end user function requests as possible, which means that the revenue of resource allocation may be improved. This performance is measured by the acceptance rate of user function requests. The calculation method is,
\begin{equation}
ACR=\lim_{T \to \infty}\frac{\sum\limits_{t=0}^{T}acc\_num(G^R,t)}{\sum\limits_{t=0}^{T}arr\_num(G^R,t)},
\end{equation}
where $acc\_num(G^R,t)$ represents the number of user function requests successfully mapped to SAGIN, and $arr\_num(G^R,t)$ represents the total number of resource requests sent by users in the resource allocation phase.

\subsubsection{Optimization Objective}

The long-term average revenue of network resource allocation, long-term revenue rate and user function request acceptance rate are jointly optimized. The purpose is to maximize the value after joint optimization, which is defined as,
\begin{equation}
maximize \,\, O=O_1+O_2+O_3.
\label{opt}
\end{equation}
\textit{where}
\begin{equation}\nonumber
O_1=AR, \, O_2=R/C, \, O_3=ACR.
\end{equation}

\subsection{Markov Decision Model}

We use MDP to build a model for the resource allocation process of SAGIN. In the SAGIN resource allocation scenario, we use a custom neural network as the agent, and deploy it on each edge server of different network segments, so that it can obtain the optimal resource allocation strategy through interactive training with the local physical network environment. The detailed composition of the agent will be introduced below. The MDP is represented by a quadruple $<\mathcal{S},\mathcal{A},\mathcal{P},\mathcal{R}>$, where $\mathcal{S}=\{\mathcal{S}_1,\mathcal{S}_2,...,\mathcal{S}_U\}$ is the state of each edge physical domain, and $U$ is the total number of edge physical domains in SAGIN. $\mathcal{A}$ represents the joint action set of agents in each physical domain. $\mathcal{P}$ is the probability of state transition. $\mathcal{R}$ represents the reward set of all edge physical domain agents.

\subsubsection{State}

The state of the edge physical domain observed by each agent is represented by $s_i(t) \in \mathcal{S}$. In SAGIN, the environment state that the agent needs to observe mainly includes the available capacity of various network resources and the connection relationship of physical nodes. Therefore, at time $t$, the environmental state observed by the agent in edge physical domain of number $i$ is expressed as,
\begin{equation}
s_i(t)=\{CPU(N_i^P),STO(N_i^P),BW(L_i^P),C_i(n^p)\},
\end{equation}
where $C_i(n^p)$ represents the set of physical nodes directly connected to the node $n^p$ in the edge physical domain.

After performing an action to the current state, the environment will shift to a new action $s_i(t+1)$ according to the agent's action. It should be noted that the transition from current state $s_i(t)$ to next state $s_i(t+1)$ is jointly determined by all $U$ agents.

\subsubsection{Action}

The action performed by the agent in edge physical domain of number $i$ is denoted as $a_i(t) \in \mathcal{A}_i$. In SAGIN, it represents the specific network resource allocation strategy adopted by the agent, i.e., which physical node is the end user function request node mapped to, and which physical link is the request link mapped to.

\subsubsection{Reward}

The instant reward obtained by the action $a_i(t)$ performed by the agent in edge physical domain of number $i$ is denoted as $r_i(t) \in \mathcal{R}_i$. Applying it to SAGIN, the reward represents the joint optimization result of the long-term average revenue of network resource allocation, the long-term revenue rate and the acceptance rate of user function requests, specifically expressed as,
\begin{equation}
\begin{aligned}
r_i(t)=O=O_1+O_2+O_3 = AR+R/C+ACR \\
 = \lim_{T \to \infty}\frac{\sum\limits_{t=0}^{T}Revenue(G^R,t)}{T} + \\
\lim_{T \to \infty}\frac{\sum\limits_{t=0}^{T}Revenue(G^R,t)}{\sum\limits_{t=0}^{T}Cost(G^R,t)} + \\
\lim_{T \to \infty}\frac{\sum\limits_{t=0}^{T}acc\_num(G^R,t)}{\sum\limits_{t=0}^{T}arr\_num(G^R,t)}.
\label{nopt}
\end{aligned}
\end{equation}

Strategy $\pi_i(s_i,a_i)$ is defined as mapping action $a_i$ to probability $\mathcal{P}(a_t=a|s_t=s)$ under state $s_i$. In distributed SAGIN, the resource allocation strategy adopted by $U$ agents is defined as a joint strategy vector $\pi=(\pi_1(s_1),\pi_2(s_2),...,\pi_U(s_U))$. Based on the probability strategy, the joint optimization objective defined by formula (\ref{nopt}) can be transformed into the following form,
\begin{equation}
\mathcal{V}_i(s_i,\pi)=E[\sum\limits_{\tau=0}^{+\infty}\gamma^\tau r_i(t+\tau)|s_i(t)=s_i,\pi],
\end{equation}
where $E$ is the expected operation. $\gamma \in [0,1]$ is a discount factor. $\tau$ is the time slot index starting from the time slot $t$.

At each time $t$, the agent independently selects the resource allocation strategy in the state $s_i(t)$ to maximize its optimization goal $\mathcal{V}_i(s_i,\pi)$, and then obtains the individual agent reward $\pi_i$ based on the joint optimization result. The ultimate goal of the agent in each edge physical domain is to learn the optimal strategy from state $s_i(t)$, which is obtained by continuous iterative training.

\section{Distributed DRL Network Resource Management Algorithm}\label{part4}

\subsection{DRL Algorithm Framework}

\begin{figure}[!h]
\centering
\includegraphics[width=1\columnwidth]{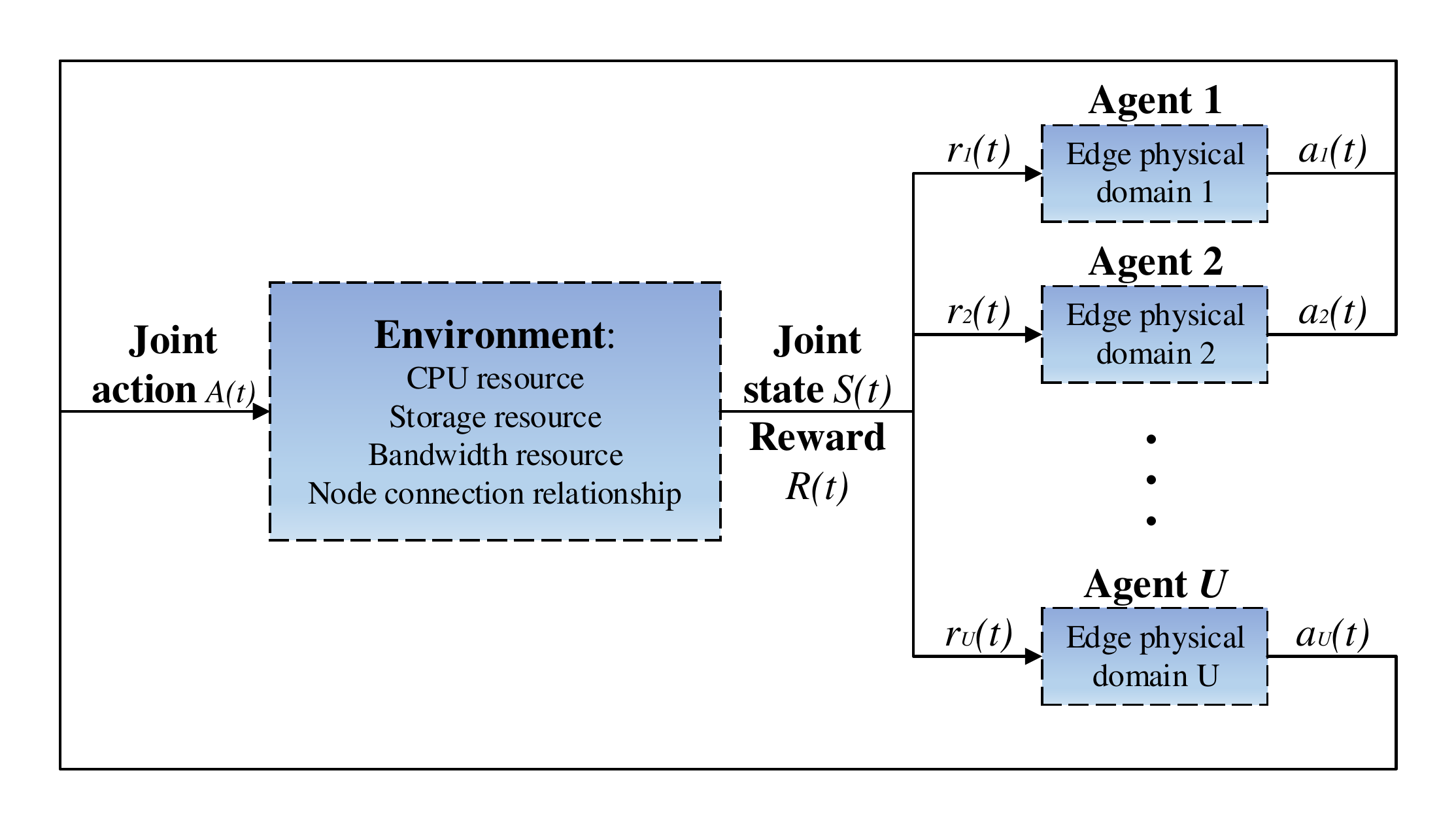}
\caption{Deep reinforcement learning framework based on distributed architecture. Intelligent agents are deployed in each edge physical domain, and each intelligent agent cooperates and interacts with the environment.}
\label{fig_2}
\end{figure}

The DRL algorithm framework used in this paper is shown in Fig. \ref{fig_2}. Each edge physical domain runs DRL algorithm independently. At time $t$, for the $i$-th edge physical domain in state $s_i(t)$, the action taken by the intelligent agent is $a_i(t) \in \pi_i(s_i)$. At the current time, the joint state of all edge physical domains is $\mathcal{S}(t)=\{s_1(t),s_2(t),...,s_U(t)\}$, and the joint action of all agents in SAGIN is $\mathcal{A}(t)=\{a_1(t),a_2(t),...,a_U(t)\}$. Then the joint state of the edge physical domain is transferred to a new joint state $\mathcal{S}(t+1)=\{s_1(t+1),s_2(t+1),...,s_U(t+1)\}$. At this time, the agent in each edge physical domain gets a reward $r_i(t)$. Since the transition between the new state $\mathcal{S}(t+1)$ and the starting state $\mathcal{S}(t)$ only depends on the action $\mathcal{A}(t)$, this is a typical MDP.

\subsection{Feature Extraction and Intelligent Agent}

Taking an edge physical domain as an example, other edge physical domains adopt the same physical network feature extraction method. The purpose of extracting physical network features is to build a training environment for intelligent agents. As mentioned in section \ref{part3}-$A$, we set three main network attributes for SAGIN: computing resource, storage resource and bandwidth resource, so these three main resource attributes are to be extracted. It should be noted that the above three network attributes can only reflect the local attribute characteristics of a physical node or a physical link independently, which is not comprehensive. Therefore, we consider extracting another attribute that can reflect the characteristics of the global network, which is called the average distance to other physical nodes in the physical domain, which is defined as,
\begin{equation}
AD(n^p)=\frac{\sum\limits_{k=1}^{|N^P|-1}dst(n^p,n_k^p)}{|N^P|},
\end{equation}
where $dst(n^p,n_k^p)$ represents the distance from the physical node $n^p$ to other physical nodes, which is measured by the number of hops between physical nodes. $|N^P|$ represents the number of physical nodes in the edge physical domain.

CPU resource is the main resource in network environment and any host cannot perform core computing tasks without CPU resource. In SAGIN, especially for the air nodes such as satellites or UAVs, their storage resources are often limited due to their size. Therefore, it is particularly important to arrange the storage resources reasonably for SAGIN. Bandwidth resource is the main network resource used in link communication. Long distance communication such as space-air and air-ground requires sufficient bandwidth resource. The optimal link can be selected by calculating the average distance to other physical nodes to map the request link, so as to save bandwidth resources and reduce resource allocation cost to a certain extent. Therefore, it is reasonable and practical to extract the above four network resources for intelligent agent.

Taking the $k$-th physical node $n_k^p$ as an example, the extracted feature vector is,
\begin{equation}
v(n_k^p)=[CPU(n_k^p),STO(n_k^p),BW(n_k^p),AD(n_k^p)],
\end{equation}
where $BW(n_k^p)$ represents the sum of bandwidth connected to the physical node $n_k^p$.

The feature vectors of all physical nodes in the edge physical domain form a feature matrix, which is expressed as,
\begin{equation}
\begin{aligned}
FM=
\begin{bmatrix}
CPU(n_1^p) & STO(n_1^p) & BW(n_1^p) & AD(n_1^p) \\
... & ... & ... & ... \\
CPU(n_k^p) & STO(n_k^p) & BW(n_k^p) & AD(n_k^p) \\
... & ... & ... & ... \\
CPU(n_u^p) & STO(n_u^p) & BW(n_u^p) & AD(n_u^p)
\end{bmatrix}
,
\end{aligned}
\end{equation}
where $u$ is the number of physical nodes in the edge physical domain. When new end user function requests arrive at the physical domain, the intelligent agent will extract a feature matrix from the physical network as its input, which not only ensures the dynamic of physical resources, but also guarantees that the intelligent agent can train in the environment close to the real network.

The essence of the intelligent agent is a four-layer neural network. The number of neurons in the input layer is determined by the number of physical nodes in each edge physical domain. Perform convolution operations on each feature vector in the convolution layer to calculate the available resources of each feature vector. The calculation method is,
\begin{equation}
v_{k}'=\omega \cdot v_k + d,
\end{equation}
where $\omega$ represents the weight vector of the convolution kernel, and $d$ represents the deviation.

In the fully connected layer, the softmax operation is performed on each feature vector after the convolution operation to obtain the probability distribution of each physical node. The sum of the mapped probabilities of all physical nodes in each edge physical domain is 1. The result represents the probability of being selected by the requested node. The calculation method is,
\begin{equation}
p_k=e^{v_{k}'} \cdot \frac{1}{\sum\limits_{i=1}^u e^{v_{i}'}}.
\label{pro}
\end{equation}

Finally, through the output layer, a group of candidate physical nodes with mapping probability in each edge physical domain is obtained.

\subsection{A General SAGIN Resource Management Framework based on distributed DRL}

\begin{figure*}[!h]
\centering
\includegraphics[width=1\textwidth]{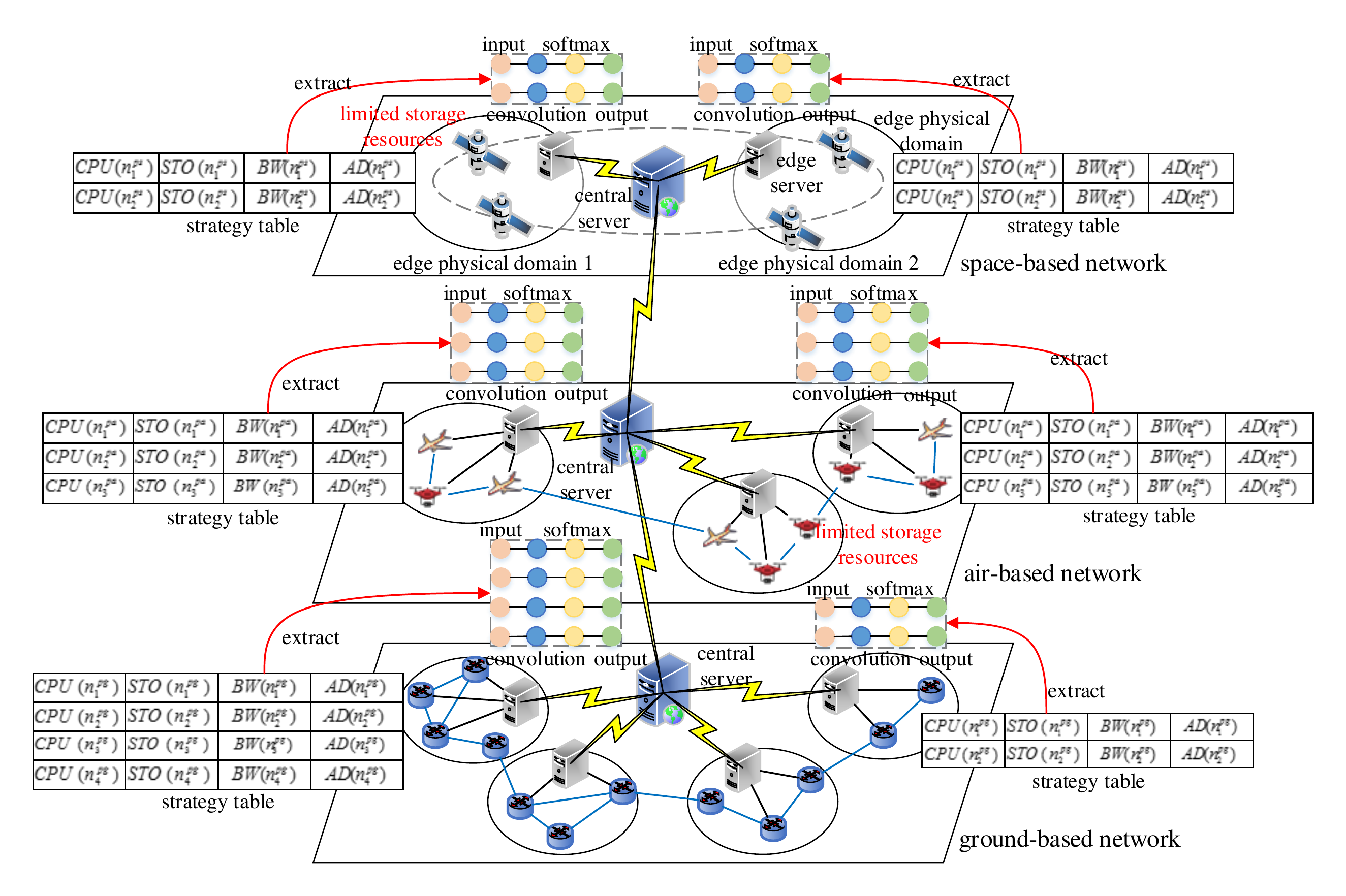}
\caption{A general SAGIN resource management architecture based on distributed DRL. The strategy table records the resource information in each physical domain. When a user function request arrives at the SAGIN, the intelligent agent will extract a feature matrix from the underlying network as input.}
\label{fig_3}
\end{figure*}

We design a distributed DRL framework as shown in Fig. \ref{fig_3}. Different network segments of SAGIN may be divided into one or more edge physical domains, and an edge server is deployed in each edge physical domain, which is the main place for the implementation of DRL algorithm. In order to collect resource information in the physical domain, each edge server is also equipped with a strategy table. The content of the strategy table is consistent with the content of the feature matrix. Whenever a user function request arrives, the agent will extract a feature matrix from the physical domain, and at the same time record the corresponding resource attribute information in the strategy table of the physical domain. Therefore, the content of the strategy table also changes dynamically.

In addition to the edge servers in each edge physical domain, each network segment also deploys a central server to summarize the resource information and network model of each edge server in the network segment. It should be noted that in order to avoid channel congestion caused by uploading too much information, only the model parameters and strategy tables of each edge physical domain need to be uploaded. The central servers of different network segments can communicate with each other for collaborative management and share the latest resource management information.

\subsection{Training and Testing}

We use a four-layer neural network as a DRL agent to participate in the training. The training environment is the feature matrix extracted from the physical network. The action performed by the agent is to allocate network resources for the end user function request based on the physical node mapping probability and the Floyd algorithm. We use the joint optimization result calculated by formula (\ref{nopt}) as the agent's reward. Then we use the gradient descent method to train the neural network, the gradient calculation method is,
\begin{equation}
g=\alpha \cdot O \cdot g',
\label{gra}
\end{equation}
where $\alpha$ is the training rate of the agent, which can control the training speed and the size of the training gradient. $O$ is the result of joint optimization, i.e., the immediate reward of the agent. $g'$ is the stacking gradient. We use a batch update strategy for stacked gradients. The reason for this is to speed up the entire training process on the one hand, and to obtain a more stable gradient on the other hand.

The training process of SAGIN storage resources management algorithm based on distributed DRL is shown in Algorithm \ref{train}. Line 4 and line 5 calculate the mapping probability of each physical node and training gradient respectively. Line 6 represents the allocation of node resources, and line 7 represents the allocation of link resources. In the link resource allocation stage, we use the breadth-first search (BFS) strategy to explore the optimal link. The purpose of training is to improve the usability of the network model, so the final training output is a set of high-quality neural network parameters.

\begin{algorithm}
  \caption{Training process of SAGIN resource management algorithm based on distributed DRL}
  \label{train}
  \begin{algorithmic}[1]
    \Require
        {$G^P,\,\,G^R\,\,from\,\,training\,\,set,\,\,number\,\,of\,\,iterations$};
    \Ensure
        {$neural\,\,network\,\,parameters$};
    \While {$iteration<num$}
    \For {$n^r \in N^R$}
    \State {$get\,\,FM$};
    \State {$calculate\,\,probability\,\,by\,\,(\ref{pro})$};
    \State {$calculate\,\,gradient\,\,by\,\,(\ref{gra})$};
    \If {$node\,\,resource\,\,allocation\,\,completed$}
    \State {$use\,\,BFS\,\,complete\,\,link\,\,resource\,\,allocation$};
    \EndIf
    \If {$node\,\,and\,\,link\,\,resource\,\,allocation\,\,completed$}
    \State {$calculate\,\,reward\,\,by\,\,(\ref{nopt})$};
    \EndIf
    \State {$apply\,\,gradient\,\,update\,\,parameters$};
    \State {$iteration++$};
    \EndFor
    \EndWhile
  \end{algorithmic}
\end{algorithm}

The testing process of SAGIN storage resources management algorithm based on distributed DRL is shown in Algorithm \ref{test}. We use the trained neural network to obtain the mapping probability, and then use the greedy strategy to complete the entire resource allocation process according to the probability. Line 2 represents the allocation of node resources, and line 3 indicates that the BFS strategy is directly used to complete the allocation of link resources.

\begin{algorithm}
  \caption{Testing process of SAGIN resource management algorithm based on distributed DRL}
  \label{test}
  \begin{algorithmic}[1]
    \Require
        {$G^P,\,\,G^R\,\,from\,\,test\,\,set$};
    \Ensure
        {$revenue,\,\,revenue\,\,ratio,\,\,and\,\,acceptance\,\,rate$};
    \For {$n^r \in N^R$}
    \State {$node\,\,resource\,\,allocation\,\,based\,\,on\,\,probability$};
    \State {$link\,\,resource\,\,allocation\,\,based\,\,on\,\,BFS$};
    \If {$node\,\,and\,\,link\,\,resource\,\,allocation\,\,completed$}
    \State $return\,\,(SUCCESS)$;
    \EndIf
    \EndFor
  \end{algorithmic}
\end{algorithm}

The training process of the algorithm is performed offline, and the testing process is performed online. Assume that the total number of nodes in SAGIN is $N$ and the total number of links is $E$. In the node resource allocation stage, network resource allocation is performed according to the physical node mapping probability, and the time complexity of this process is $O(N)$. In the link resource allocation stage, the BFS strategy is adopted, and the time complexity is $O(E)$. Therefore, the total time complexity of the algorithm is $O(N+E)$.

\section{Performance Evaluation}\label{part5}

\subsection{Simulation Environment and Parameter Setting}

The simulation experiment is carried out in Pycharm2018.2 + anaconda3-5.2.0. We use CodeBlocks16.01 programming to generate text documents to save relevant network information. The topology dynamics of SAGIN will increase the difficulty of research. Since the topology of SAGIN is relatively fixed in some time period, we carry out research on the fixed topological structure in a certain time period. We generate a medium-sized physical network through programming to simulate SAGIN, and divide it into three network segments: space-based network, air-based network and ground-based network. All information of the physical network is stored in a text file. The difference between different network segments is mainly reflected in the number of physical nodes and resource capacity. In addition, we have also generated 2,000 text files through programming to save the relevant information of end user function requests, including the number of requested nodes, link connection relationships, and resource requirements. Among them, 1,000 simulate the end user function requests in the training stage, and another 1,000 simulate the end user function requests in the test stage.

TABLE \ref{tab_2} summarizes the main parameters of the simulation experiment.

\begin{table}
\centering
\caption{Simulation Parameter Setting}
\renewcommand\arraystretch{1.5}
\begin{tabular}{p{62mm} p{19mm}}
\hline
\hline
Parameter &  Value \\
\hline
IDs of different network segments & \{0,1,2\} \\
number of ground nodes & 60 \\
number of air nodes & 30 \\
number of space nodes & 10 \\
CPU resource capacity of ground node & U[50,100]Mbps \\
storage resource capacity of ground node & U[50,100]MB \\
CPU resource capacity of air node & U[50,80]Mbps \\
storage resource capacity of air node & U[50,80]MB \\
CPU resource capacity of space node & U[50,80]Mbps \\
storage resource capacity of space node & U[50,80]MB \\
ground link bandwidth resource capacity & U[50,100]MB/s \\
air link bandwidth resource capacity & U[50,80]MB/s \\
space link bandwidth resource capacity & U[50,80]MB/s \\
number of end user function request node & U[2,10] \\
CPU requirement of end user function request node & U[1,50]Mbps \\
storage requirement of user function request node & U[1,50]MB \\
bandwidth requirement of user function request link & U[1,50]MB/s \\
request link connection probability & 50\% \\
initial learning rate & 0.005 \\
\hline
\hline
\end{tabular}
\label{tab_2}
\end{table}

\subsection{Comparison Algorithms}

We compare the distributed DRL based network resource allocation (DDRL-NRA) algorithm with the other two general network resource algorithms. In order to ensure the fairness of the contrast, we set up the same experimental environment as the algorithm in this paper.

The first algorithm is the RLVNE algorithm proposed in \cite{14}, which is a virtual network embedding algorithm based on RL. Since the essence of virtual network embedding is the process of network resource allocation, it is also a general network resource allocation algorithm. The main idea is to use a RL agent called a policy network to derive node mapping probabilities, and use historical data requested by user functions to train the RL agent. Finally, based on BFS strategy, bandwidth resources are allocated for the user request link.

The second algorithm is NRMVNE algorithm proposed in reference \cite{z4}, which is also a heuristic based general network resource allocation algorithm. The algorithm divides the resource allocation process into two stages. The first stage is to allocate network resources for user function request nodes, and the second stage is to allocate network resources for user function request links. In the first stage, the resource metrics of each physical node and request node are calculated, and the physical nodes and request nodes are arranged according to the value. After that, the request nodes are mapped to the physical nodes in turn. In the second stage, the request links are sorted according to the bandwidth requirement, and then the shortest path algorithm is used to allocate bandwidth resources for the request links.

\subsection{Results and Analysis}

\subsubsection{Training Performance Verification}

\begin{figure*}[!h]
\centering
\includegraphics[width=1\textwidth]{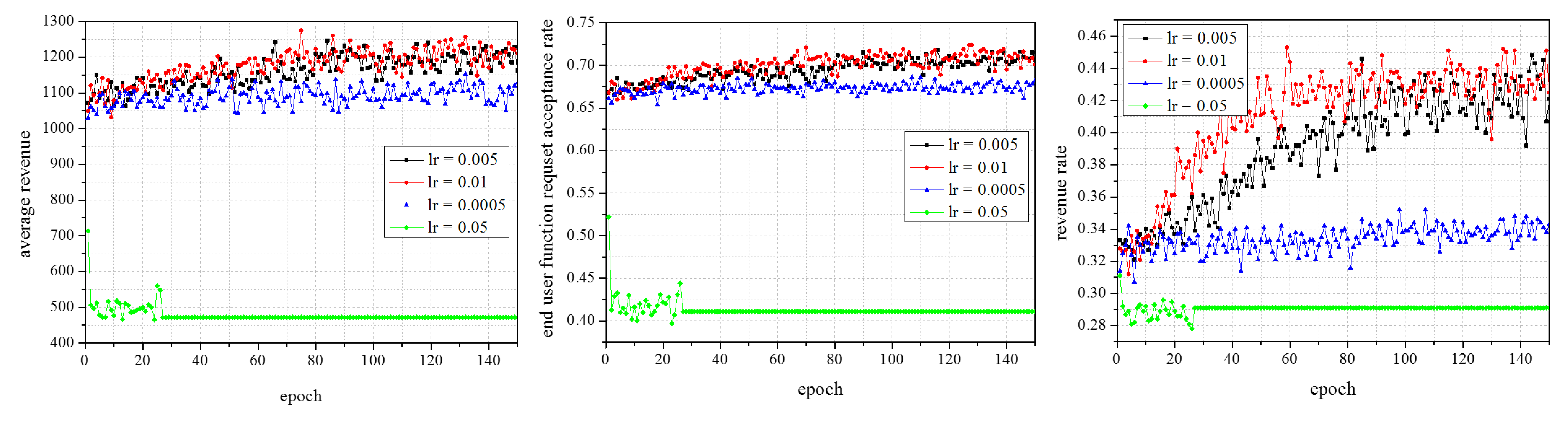}
\caption{Training results. From left to right are the long-term average revenue of network resource allocation, the acceptance rate of end user function request and the revenue rate.}
\label{fig_4}
\end{figure*}

The training of DRL is more difficult, especially when the underlying physical network is more complex. In order to obtain the best performance of the DRL agent, we first explore the convergence of the intelligent agent under different learning rates, and then we train the intelligent agent for a sufficient number of times to ensure that it can reach a stable level. Specifically, we respectively set the learning rate of the DRL agent to 0.0005, 0.005, 0.01, and 0.05, and then observe the training results of the agent from the three aspects of network resource distribution revenue, user function request acceptance rate and long-term revenue rate. As shown in Fig. \ref{fig_4}.

Different learning rates of intelligent agent will lead to significant differences in training performance. When the learning rate is 0.005 and 0.01, there is little difference in training performance. But when the learning rate is too large or too small, the performance of the algorithm is poor. In order to obtain the best training performance, we fix the training learning rate of DDRL-NRA to 0.005.

At the beginning of training, the performance of agent is the most unstable. This is because the agent has just entered into a completely unfamiliar SAGIN environment, and the network parameters are random. So at this time, from the three training indicators, the agent's performance has not reached the ideal effect. In the middle of training, the agent performance begins to show a stable trend. Because after a period of learning, the agent begins to be familiar with the surrounding environment. By continuously extracting the attributes of the network resources of SAGIN, the agent may perform an action that makes it more profitable, so the training performance at this time is better than that at the early stage of training. At the end of the training, the training performance has reached a stable level. On the one hand, the agent has been fully familiar with the network environment, and the action can make it get a larger reward signal. On the other hand, because the performance of neural network model reaches the limit, the training effect cannot be further improved at this time.

\subsubsection{Test Results and Analysis}

\begin{figure}[!h]
\centering
\includegraphics[width=1\columnwidth]{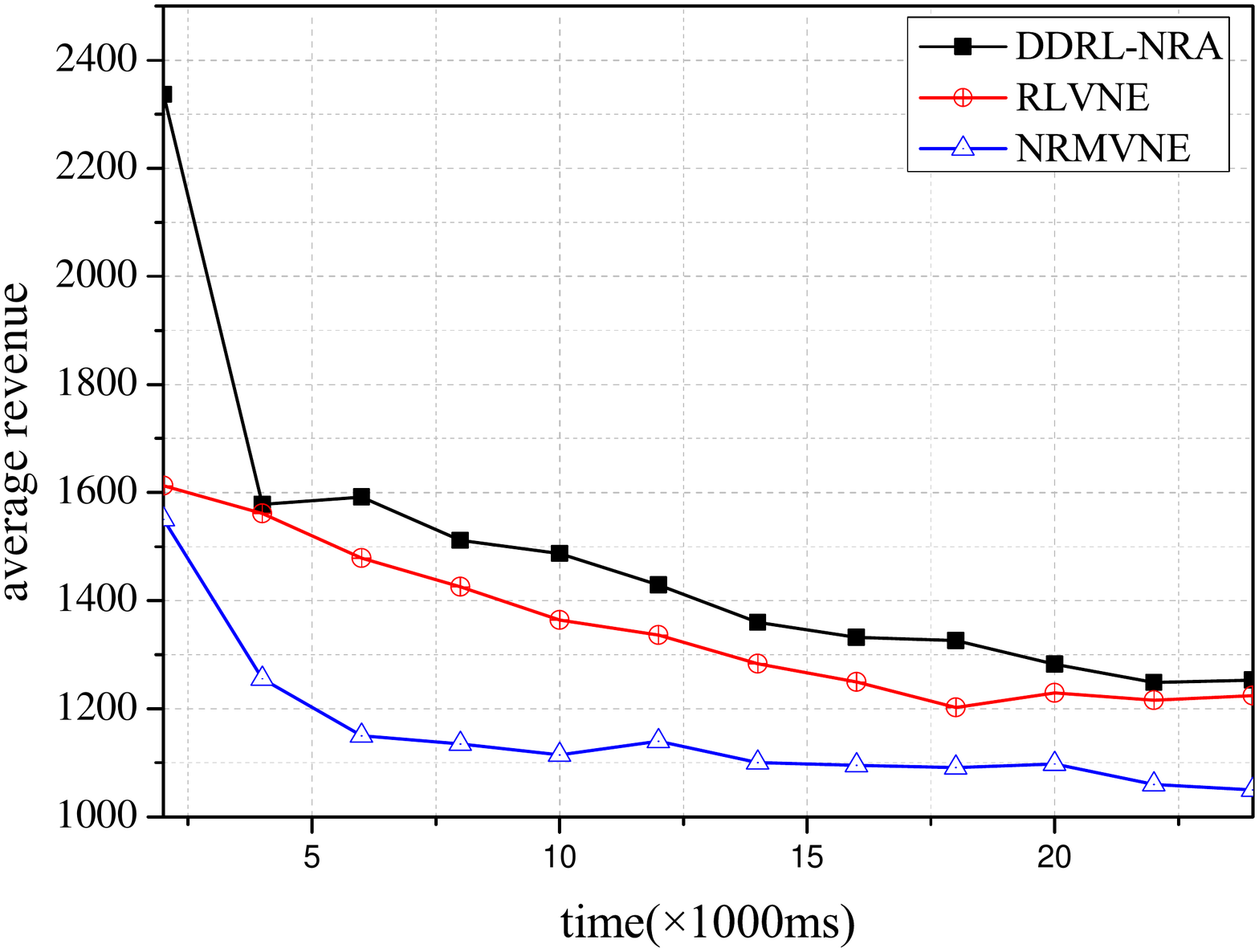}
\caption{Test result of long term average revenue of network resource allocation.}
\label{fig_5}
\end{figure}

\begin{figure}[!h]
\centering
\includegraphics[width=1\columnwidth]{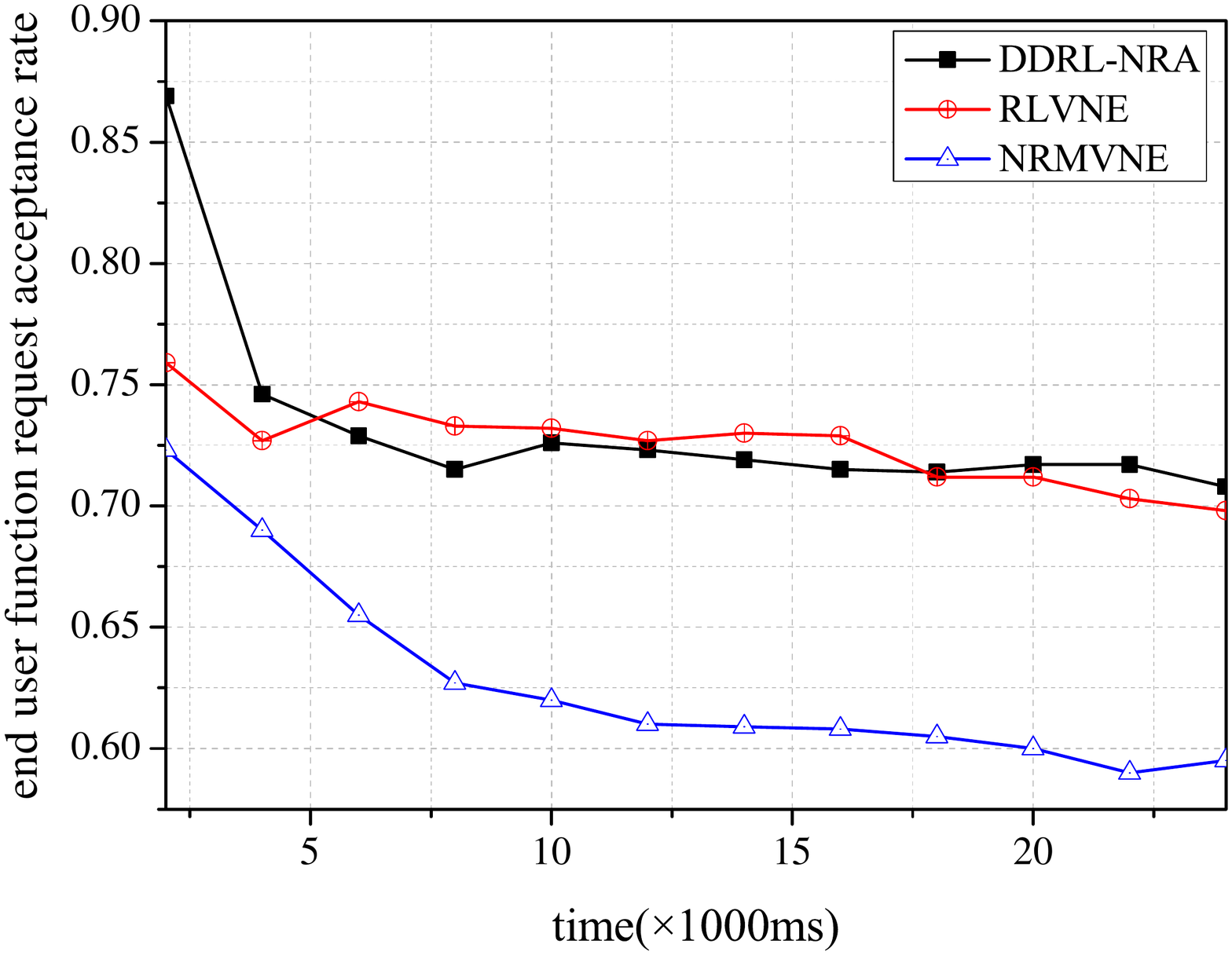}
\caption{Test result of the acceptance rate of end user requests.}
\label{fig_6}
\end{figure}

We test the algorithm with the remaining 1,000 end user function request files as the test set. Fig. \ref{fig_5} reflects the test results of the DDRL-NRA algorithm in terms of the long-term average revenue of resource allocation. Judging from the overall trend of the three algorithms, the performance curve shows a gradual downward trend. This is because the long-term average revenue of resource allocation is determined by the remaining capacity of physical network resources. In the early stage of the arrival of user function requests, SAGIN resources are relatively abundant. At this time, more end user function requests can be received, so network operators can obtain higher revenue. However, with the development of resource allocation, the number of available resources in each edge physical domain is decreasing. At this time, the resource capacity cannot meet most end user function requests. Therefore, the number of end user function requests for successful network resources is reduced, and the revenue of network operators are also beginning to decrease. This also illustrates the reason why the acceptance rate of end user function requests shown in Fig. \ref{fig_6} is reduced.

Compared with other two algorithms, the DDRL-NRA algorithm has obvious advantages in terms of revenue and acceptance rate. For large-scale SAGIN scenario, the use of a distributed management solution is more effective than a centralized management solution. We have deployed DRL on each edge server, and intelligent agent can execute resource allocation algorithms in each edge physical domain without having to concentrate detailed resource allocation information on the central server. RLVNE is a centralized network resource allocation algorithm. It uses historical data to train RL agent. It does not consider network resources as a dynamic process and ignores the real-time changes of the underlying network. NRMVNE is a heuristic algorithm. All resource allocation decision rules are made manually. The experimental results show that the network resource allocation algorithm based on ML is better than the heuristic-based network resource allocation algorithm.

\begin{figure}[!h]
\centering
\includegraphics[width=1\columnwidth]{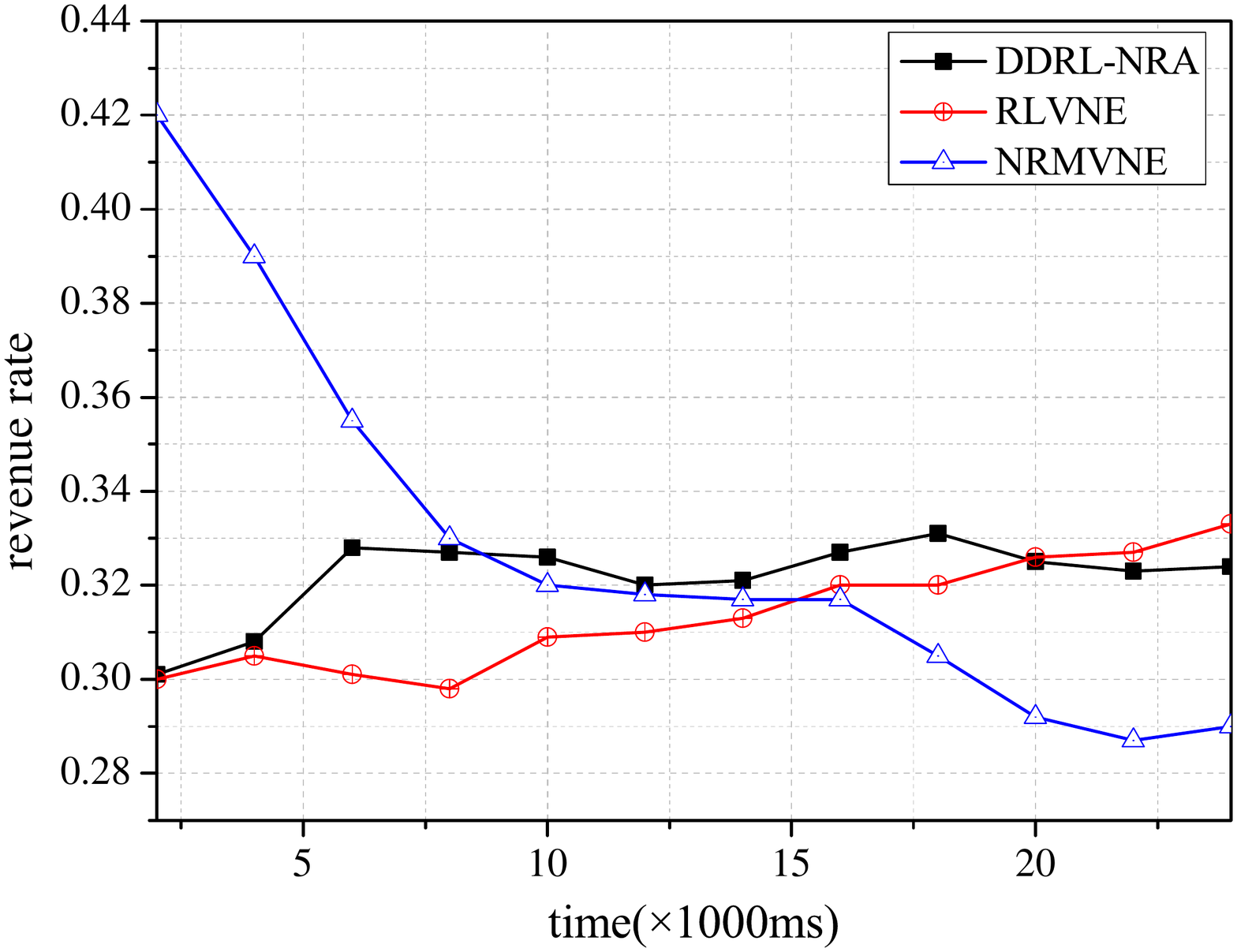}
\caption{Test result of revenue rate.}
\label{fig_7}
\end{figure}

Fig. \ref{fig_7} shows the test results of the algorithm in terms of revenue rate. From the results, the overall effect of the DDRL-NRA algorithm is the best. The revenue rate of network resource allocation is determined by the revenue and cost of resource distribution. It does not only depend on the number of network resources. From the perspective of the revenue of network resource allocation, the DDRL-NRA algorithm can obtain higher network revenue. At the beginning, the NRMVNE algorithm achieved a higher revenue rate. This is because the algorithm prioritizes the allocation of nodes and links with sufficient resources without considering the rationality of the coordinated allocation of multiple resources. Although the algorithm can achieve higher revenue in the early stage, with the rapid consumption of resources, the revenue rate of the algorithm in the later stage also decreases rapidly. Therefore, when the three algorithms are compared in terms of revenue rate, the DDRL-NRA algorithm has a slight advantage.

\begin{figure}[!h]
\centering
\includegraphics[width=1\columnwidth]{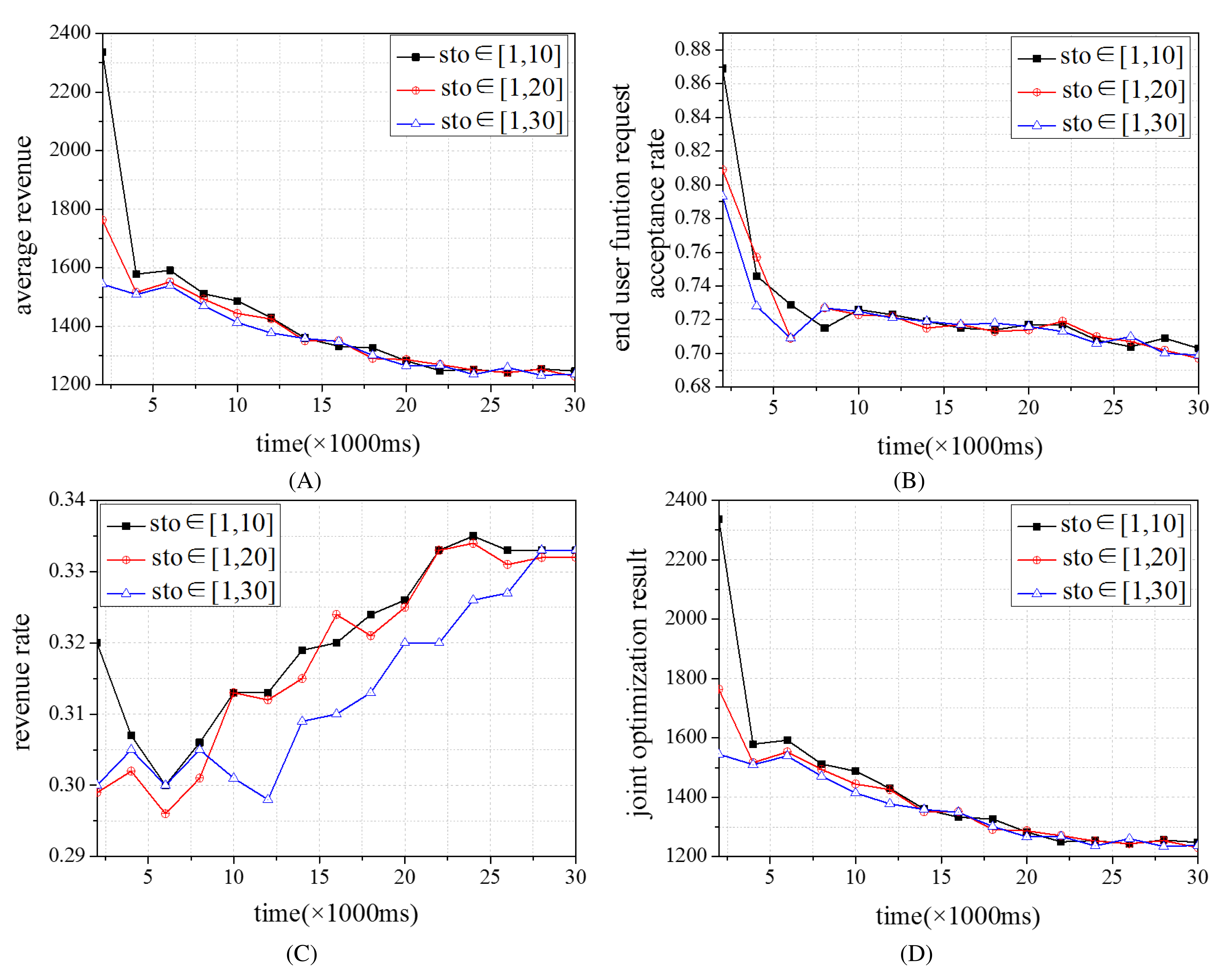}
\caption{Proposed algorithm flexibility test. Under different storage resource requirements: (A) the long-term average revenue of network resource allocation, (B) the acceptance rate of end-user function requests, (C) the revenue rate, (D) the result of joint optimization.}
\label{fig_8}
\end{figure}

In addition to verifying the effectiveness of the DDRL-NRA algorithm, we verify the flexibility of the algorithm by changing the resource requirements of end user function requests. Specifically, we adjust the storage resource requirements of the initial end user function requests to [1,20], [1,10], and then conduct experiments from three aspects: resource allocation revenue, end user request acceptance rate and revenue rate. Finally, we also show the comparison of the joint optimization results under three different storage resource requirements. The experimental results are shown in Fig. \ref{fig_8}.

From the experimental results, the three performance indicators of the algorithm show similar trend under three different resource requirements, which shows that the change of resource demand will not affect the effectiveness of the algorithm. One obvious feature is that when the storage resource demand of user function request is between [1,10], the best experimental effect can be obtained. This is because after reducing the storage resource demand, SAGIN can receive more end user function requests, so the above three performance indicators have risen. In the later stage of the experiment, the joint optimization result under the three storage resource requirements have little difference. This is because the underlying network resources are not much remaining after a period of time consumption, and only a small number of user function requests can be allocated network resources. The experimental results further illustrate that the algorithm decision can be flexibly changed as the network environment changes, which intuitively reflects the effectiveness of the algorithm for storage resource management. Therefore, the DDRL-NRA algorithm has good flexibility. The joint optimization results are determined by the revenue of network resource allocation, the acceptance rate of end user function request and the revenue rate. According to the test results of Fig. \ref{fig_5} to Fig. \ref{fig_7}, the joint optimization results of the algorithm are declining with time, which further illustrates the effectiveness of the DDRL-NRA algorithm in network resource allocation.

\section{Conclusions and Future Work}\label{part6}

Due to the heterogeneous, time-varying and self-organizing characteristics of SAGIN, the centralized resource management scheme is extremely difficult to implement. In addition, SAGIN's air nodes are easily limited by storage capacity, and how to reasonably schedule limited storage resources is the key to task execution. We model the resource allocation of SAGIN as a MDP, and then propose a SAGIN storage resource allocation algorithm based on distributed DRL. We deploy an intelligent server with DRL capabilities in each edge physical domain. The intelligent agent is composed of the basic elements of the neural network, and it is trained in an environment built by SAGIN resource attributes. Through training, the optimal resource allocation decision can be made for each end user function request. In the experimental simulation stage, we verify the performance of the proposed algorithm from two perspectives of training and testing. In addition, the algorithm can also flexibly allocate network resources according to changes in end user function requests. In general, the algorithm we proposed has achieved an ideal resource allocation effect.

As part of our future work, we will consider the impact of other resource elements on SAGIN, such as battery, energy and spectrum resources. In addition, we will try to adjust the neural network model or adopt other AI technologies to carry out more comprehensive and effective resource management of SAGIN. At the same time, it needs to be considered that SAGIN is a large-scale complex network, so it is necessary to design a more complex and deeper neural network model to improve its ability to deal with complex problems. SAGIN is a time-varying network, and the underlying network topology and resource conditions are constantly changing. It will be more practical to study the problem of network resource allocation under dynamic topology. Finally, we will explore the scalability of the algorithm and its theoretical effects in other distributed application scenarios.

\ifCLASSOPTIONcaptionsoff
  \newpage
\fi


\begin{thebibliography}{1}
\bibitem{ed1}
L. Liu, C. Chen, Q. Pei, S. Maharjan and Y. Zhang, ``Vehicular Edge Computing and Networking: A Survey,'' {\em Mobile Networks and Applications}, vol. 26, no. 3, pp. 1145-1168, 2021.

\bibitem{l1}
C. Qiu, X. Wang, H. Yao, J. Du, F. R. Yu and S. Guo, ``Networking Integrated Cloud–Edge–End in IoT: A Blockchain-Assisted Collective Q-Learning Approach,'' {\em IEEE Internet of Things Journal}, vol. 8, no. 16, pp. 12694-12704, Aug. 2021.

\bibitem{j7}
J. Wang, C. Jiang, L. Kuang and B. Yang, ``Iterative Doppler Frequency Offset Estimation in Satellite High-Mobility Communications,'' {\em IEEE Journal on Selected Areas in Communications}, vol. 38, no. 12, pp. 2875-2888, Dec. 2020.

\bibitem{j6}
J. Du, C. Jiang, J. Wang, Y. Ren and M. Debbah, ``Machine Learning for 6G Wireless Networks: Carrying Forward Enhanced Bandwidth, Massive Access, and Ultrareliable/Low-Latency Service,'' {\em IEEE Vehicular Technology Magazine}, vol. 15, no. 4, pp. 122-134, Dec. 2020.

\bibitem{q1}
C. Qiu, H. Yao, X. Wang, N. Zhang, F. R. Yu and D. Niyato, ``AI-Chain: Blockchain Energized Edge Intelligence for Beyond 5G Networks,'' {\em IEEE Network}, vol. 34, no. 6, pp. 62-69, 2020.

\bibitem{jcx1}
C. Jiang, Y. Chen, K. J. R. Liu and Y. Ren, ``Renewal-Theoretical Dynamic Spectrum Access in Cognitive Radio Network with Unknown Primary Behavior,'' {\em IEEE Journal on Selected Areas in Communications}, vol. 31, no. 3, pp. 406-416, 2013.

\bibitem{jcx2}
C. Jiang, Y. Chen, Y. Gao and K. J. R. Liu, ``Joint Spectrum Sensing and Access Evolutionary Game in Cognitive Radio Networks,'' {\em IEEE Transactions on Wireless Communications}, vol. 12, no. 5, pp. 2470-2483, 2013.

\bibitem{jcx3}
X. Zhu, C. Jiang, L. Kuang, N. Ge and J. Lu, ``Non-orthogonal Multiple Access Based Integrated Terrestrial-Satellite Networks,'' {\em IEEE Journal on Selected Areas in Communications}, vol. 35, no. 10, pp. 2253-2267, Oct. 2017.

\bibitem{j2}
Y. Li, H. Zhang, K. Long, C. Jiang and M. Guizani, ``Joint Resource Allocation and Trajectory Optimization with QoS in UAV-based NOMA Wireless Networks,'' {\em IEEE Transactions on Wireless Communications}, vol. 20, no. 10, pp. 6343-6355, Oct. 2021.

\bibitem{ed2}
X. Wu, J. Li, M. Xiao, P. C. Ching and H. V. Poor, ``Multi-Agent Reinforcement Learning for Cooperative Coded Caching via Homotopy Optimization,'' {\em IEEE Transactions on Wireless Communications}, vol. 20, no. 8, pp. 5258-5272, Aug. 2021.

\bibitem{l2}
J. Feng, L. Liu, Q. Pei, F. Hou, T. Yang and J. Wu, ``Service Characteristics-Oriented Joint Optimization of Radio and Computing Resource Allocation in Mobile-Edge Computing,'' {\em IEEE Internet of Things Journal}, vol. 8, no. 11, pp. 9407-9421, Jun. 2021.

\bibitem{r3}
A. Paul, I. Kamwa and G. Jóos, ``Centralized Dynamic State Estimation Using a Federation of Extended Kalman Filters With Intermittent PMU Data From Generator Terminals,'' {\em IEEE Transactions on Power Systems}, vol. 33, no. 6, pp. 6109-6119, Nov. 2018.

\bibitem{j3}
P. Zhang, C. Jiang, X. Pang and Y. Qian, ``STEC-IoT: A Security Tactic by Virtualizing Edge Computing on IoT,'' {\em IEEE Internet of Things Journal}, vol. 8, no. 4, pp. 2459-2467, Feb. 2021.

\bibitem{ed3}
J. Yue and M. Xiao, ``Coding for Distributed Fog Computing in Internet of Mobile Things,'' {\em IEEE Transactions on Mobile Computing}, vol. 20, no. 4, pp. 1337-1350, Apr. 2021.

\bibitem{zz1}
Z. Fang, S. Shen, J. Liu, W. Ni and A. Jamalipour, ``New NOMA-Based Two-Way Relay Networks,'' {\em IEEE Transactions on Vehicular Technology}, vol. 69, no. 12, pp. 15314-15324, Dec. 2020.

\bibitem{j5}
C. Jiang, N. Ge and L. Kuang, ``AI-Enabled Next-Generation Communication Networks: Intelligent Agent and AI Router,'' {\em IEEE Wireless Communications}, vol. 27, no. 6, pp. 129-133, Dec. 2020.

\bibitem{e1}
H. Shi, S. Liu, H. Wu, R. Li, S. Liu, N. Kwok and Y. Peng, ``Oscillatory Particle Swarm Optimizer,'' {\em Applied Soft Computing}, vol. 73, pp. 316–327, 2018.

\bibitem{e2}
H. Ma, S. Shen, M. Yu, Z. Yang, M. Fei and H. Zhou, ``Multi-Population Techniques in Nature Inspired Optimization Algorithms: A Comprehensive Survey,'' {\em Swarm and Evolutionary Computation}, vol. 44, pp. 365–387, 2019.

\bibitem{z4}
Y. Ye, S. Huang, M. Xiao and Z. Ma, ``Decentralized Consensus Optimization Based on Parallel Random Walk,'' {\em IEEE Communications Letters}, vol. 24, no. 2, pp. 391-395, Feb. 2020.

\bibitem{e3}
Q. Li, Z. Cao, W. Ding and Q. Li, ``A Multi-Objective Adaptive Evolutionary Algorithm to Extract Communities in Networks,'' {\em Swarm and Evolutionary Computation}, vol. 52, pp. 100629, Feb. 2020.

\bibitem{l3}
L. Liu, J. Feng, Q. Pei, Y. Ming, B. Shang and M. Dong, ``Blockchain-Enabled Secure Data Sharing Scheme in Mobile-Edge Computing: An Asynchronous Advantage Actor–Critic Learning Approach,'' {\em IEEE Internet of Things Journal}, vol. 8, no. 4, pp. 2342-2353, Feb. 2021.

\bibitem{r5}
C. Qiu, F. R. Yu, H. Yao, C. Jiang, F. Xu and C. Zhao, ``Blockchain-Based Software-Defined Industrial Internet of Things: A Dueling Deep Q-Learning Approach,'' {\em IEEE Internet of Things Journal}, vol. 6, no. 3, pp. 4627-4639, June 2019.

\bibitem{z2}
P. Zhang, C. Wang, C. Jiang and Z. Han, ``Deep Reinforcement Learning Assisted Federated Learning Algorithm for Data Management of IIoT,'' {\em IEEE Transactions on Industrial Informatics}, vol. 17, no. 12, pp. 8475-8484, Dec. 2021.

\bibitem{j4}
Y. Zhan, P. Wan, C. Jiang, X. Pan, X. Chen and S. Guo, ``Challenges and Solutions for the Satellite Tracking, Telemetry, and Command System,'' {\em IEEE Wireless Communications}, vol. 27, no. 6, pp. 12-18, Dec. 2020.

\bibitem{ed4}
Y. Ye, M. Xiao and M. Skoglund, ``Mobility-aware Content Preference Learning in Decentralized Caching Networks,'' {\em IEEE Transactions on Cognitive Communications and Networks}, Vol. 6, no. 11, pp. 62-73, Mar. 2020.

\bibitem{1}
J. Li, W. Shi, H. Wu, S. Zhang and X. Shen, ``Cost-Aware Dynamic SFC Mapping and Scheduling in SDN/NFV-Enabled Space-Air-Ground Integrated Networks for Internet of Vehicles,'' {\em IEEE Internet of Things Journal}, pp. 1-1, 2021, doi: 10.1109/JIOT.2021.3058250.

\bibitem{2}
B. Cao, J. Zhang, X. Liu, Z. Sun, W. Cao, R. M. Nowak and Z. Lv, ``Edge-Cloud Resource Scheduling in Space-Air-Ground Integrated Networks for Internet of Vehicles,'' {\em IEEE Internet of Things Journal}, pp. 1-1, 2021, doi: 10.1109/JIOT.2021.3065583.

\bibitem{3}
Y. Wang, Z. Li, Y. Chen, M. Li, X. Lyu, X. Hou and J. Wang, ``Joint Resource Allocation and UAV Trajectory Optimization for Space–Air–Ground Internet of Remote Things Networks,'' {\em IEEE Systems Journal}, pp. 1-1, 2020, doi: 10.1109/JSYST.2020.3019463.

\bibitem{4}
F. Lyu, P. Yang, H. Wu, C. Zhou, J. Ren, Y. Zhang and X. Shen, ``Service-Oriented Dynamic Resource Slicing and Optimization for Space-Air-Ground Integrated Vehicular Networks,'' {\em IEEE Transactions on Intelligent Transportation Systems}, pp. 1-1, 2021, doi: 10.1109/TITS.2021.3070542.

\bibitem{5}
Q. Chen, W. Meng, S. Han and C. Li, ``Service-Oriented Fair Resource Allocation and Auction for Civil Aircrafts Augmented Space-Air-Ground Integrated Networks,'' {\em IEEE Transactions on Vehicular Technology}, vol. 69, no. 11, pp. 13658-13672, Nov. 2020.

\bibitem{e4}
Q. Li, Z. Cao, J. Zhong and Q. Li, ``Graph Representation Learning with Encoding Edges,'' {\em Neurocomputing}, vol. 361, pp. 29–39, 2019.

\bibitem{e5}
J. Liu, X. Wang, S. Shen, G. Yue, S. Yu and M. Li, ``A Bayesian Q-learning Game for Dependable Task Offloading Against DDoS Attacks in Sensor Edge Cloud,'' {\em IEEE Internet of Things Journal}, vol. 8, no. 9, pp. 7546–7561, May 2021.

\bibitem{e6}
J. Liu, X. Wang, S. Shen, Z. Fang, S. Yu, G. Yue and M. Li, ``Intelligent Jamming Defense Using DNN Stackelberg Game in Sensor Edge Cloud,'' {\em IEEE Internet of Things Journal}, pp. 1-1, 2021, doi: 10.1109/JIOT.2021.3103196.

\bibitem{6}
G. M. S. Rahman, T. Dang and M. Ahmed, ``Deep Reinforcement Learning Based Computation Offloading and Resource Allocation for Low-Latency Fog Radio Access Networks,'' {\em Intelligent and Converged Networks}, vol. 1, no. 3, pp. 243-257, Dec. 2020.

\bibitem{7}
J. Xu, J. Wang, Q. Qi, H. Sun and B. He, ``IARA: An Intelligent Application-Aware VNF for Network Resource Allocation with Deep Learning,'' {\em 2018 15th Annual IEEE International Conference on Sensing, Communication, and Networking (SECON), Hong Kong, China}, pp. 1-3, 2018.

\bibitem{8}
F. Jiang, L. Zhang, C. Sun and Z. Yuan, ``Clustering and Resource Allocation Strategy for D2D Multicast Networks with Machine Learning Approaches,'' {\em China Communications}, vol. 18, no. 1, pp. 196-211, Jan. 2021.

\bibitem{z3}
P. Zhang, C. Wang, G. S. Aujla, N. Kumar and M. Guizani, ``IoV Scenario: Implementation of a Bandwidth Aware Algorithm in Wireless Network Communication Mode,'' {\em IEEE Transactions on Vehicular Technology}, vol. 69, no. 12, pp. 15774-15785, Dec. 2020.

\bibitem{9}
S. Yu, X. Chen, Z. Zhou, X. Gong and D. Wu, ``When Deep Reinforcement Learning Meets Federated Learning: Intelligent Multitimescale Resource Management for Multiaccess Edge Computing in 5G Ultradense Network,'' {\em IEEE Internet of Things Journal}, vol. 8, no. 4, pp. 2238-2251, Feb. 2021.

\bibitem{10}
B. Li, W. Lu, S. Liu and Z. Zhu, ``Deep-Learning-Assisted Network Orchestration for on-Demand and Cost-Effective VNF Service Chaining in Inter-DC Elastic Optical Networks,'' {\em IEEE/OSA Journal of Optical Communications and Networking}, vol. 10, no. 10, pp. D29-D41, Oct. 2018.

\bibitem{11}
Z. Allybokus, K. Avrachenkov, J. Leguay and L. Maggi, ``Multi-Path Alpha-Fair Resource Allocation at Scale in Distributed Software-Defined Networks,'' {\em IEEE Journal on Selected Areas in Communications}, vol. 36, no. 12, pp. 2655-2666, Dec. 2018.

\bibitem{12}
C. Pan, C. Yin, N. C. Beaulieu and J. Yu, ``Distributed Resource Allocation in SDCN-Based Heterogeneous Networks Utilizing Licensed and Unlicensed Bands,'' {\em IEEE Transactions on Wireless Communications}, vol. 17, no. 2, pp. 711-721, Feb. 2018.

\bibitem{13}
H. Halabian, ``Distributed Resource Allocation Optimization in 5G Virtualized Networks,'' {\em IEEE Journal on Selected Areas in Communications}, vol. 37, no. 3, pp. 627-642, Mar. 2019.

\bibitem{14}
H. Yao, X. Chen, M. Li, P. Zhang and L. Wang, ``A Novel Reinforcement Learning Algorithm for Virtual Network Embedding,'' {\em Neurocomputing}, vol. 284, pp. 1-9, 2018.

\end{thebibliography}
\end{document}